\documentclass[a4paper,reqno]{amsart}
\pdfoutput=1
\usepackage[english]{babel}
\usepackage[linktocpage=true]{hyperref}
\usepackage{amsmath}
\usepackage{amssymb}
\usepackage{mathtools}
\usepackage[indent]{parskip}

\makeatletter
\DeclareFontFamily{OMX}{MnSymbolE}{}
\DeclareSymbolFont{MnLargeSymbols}{OMX}{MnSymbolE}{m}{n}
\SetSymbolFont{MnLargeSymbols}{bold}{OMX}{MnSymbolE}{b}{n}
\DeclareFontShape{OMX}{MnSymbolE}{m}{n}{
    <-6>  MnSymbolE5
   <6-7>  MnSymbolE6
   <7-8>  MnSymbolE7
   <8-9>  MnSymbolE8
   <9-10> MnSymbolE9
  <10-12> MnSymbolE10
  <12->   MnSymbolE12
}{}
\DeclareFontShape{OMX}{MnSymbolE}{b}{n}{
    <-6>  MnSymbolE-Bold5
   <6-7>  MnSymbolE-Bold6
   <7-8>  MnSymbolE-Bold7
   <8-9>  MnSymbolE-Bold8
   <9-10> MnSymbolE-Bold9
  <10-12> MnSymbolE-Bold10
  <12->   MnSymbolE-Bold12
}{}

\let\llangle\@undefined
\let\rrangle\@undefined
\DeclareMathDelimiter{\llangle}{\mathopen}%
                     {MnLargeSymbols}{'164}{MnLargeSymbols}{'164}
\DeclareMathDelimiter{\rrangle}{\mathclose}%
                     {MnLargeSymbols}{'171}{MnLargeSymbols}{'171}
\makeatother

\numberwithin{equation}{section}

\DeclareMathOperator{\mathd}{d}
\DeclareMathOperator{\mathe}{e}
\DeclareMathOperator{\mathi}{i}
\DeclareMathOperator{\Aut}{Aut}

\DeclareMathOperator{\Tr}{Tr}

\DeclareMathOperator{\BarnesG}{BarnesG}
\DeclareMathOperator{\AiryAi}{Ai}
\DeclareMathOperator{\AiryBi}{Bi}
\DeclareMathOperator{\erf}{erf}

\newcommand{\p}{\mathsf{p}}
\newcommand{\ak}{\mathsf{a}}
\newcommand{\schur}{\mathsf{Schur}}

\begin{document}

% Preprint
\noindent{\hfill\small\scshape{UUITP-09/21}}\par\smallskip\vskip24pt

\title{Virasoro constraints revisited}

\author{Luca Cassia}
\address{Department of Physics and Astronomy, Uppsala University,
Box 516, SE-75120 Uppsala, Sweden.}
\email[L.~Cassia]{luca.cassia@physics.uu.se}
\author{Rebecca Lodin}
\address{Department of Physics and Astronomy, Uppsala University,
Box 516, SE-75120 Uppsala, Sweden.}
\email[R.~Lodin]{rebecca.lodin@physics.uu.se}
\author{Maxim Zabzine}
\address{Department of Physics and Astronomy, Uppsala University,
Box 516, SE-75120 Uppsala, Sweden.}
\email[M.~Zabzine]{maxim.zabzine@physics.uu.se}

\thanks{All authors are supported in part by the grant ``Geometry and Physics'' from the Knut and Alice Wallenberg foundation.}

\begin{abstract}
We revisit the Virasoro constraints and explore the relation to the Hirota bilinear equations. We furthermore investigate and provide the solution to non-homogeneous Virasoro constraints, namely those coming from matrix models whose domain of integration has boundaries. In particular, we provide the example of Hermitean matrices with positive eigenvalues in which case one can find a solution by induction on the rank of the matrix model.
\end{abstract}

\maketitle

\tableofcontents

\section{Introduction}

Matrix models have established their physical significance through their appearance in the description of quantum gravity in two dimensions. They appear when transforming the integral over all possible geometries and topologies to its discrete analogue, in other words the summation over random triangulations of surfaces of arbitrary genus (as reviewed in \cite{Ginsparg:1993is}). As such they give a powerful tool to probe both simple models of 2d quantum gravity as well as the theory of strings propagating in low dimensional spacetimes.

The simplest example of a matrix model is given by the Hermitean 1-matrix model. This model can be shown to satisfy Virasoro constraints, which can be derived as Ward identities corresponding to the fact that the matrix integral is invariant under certain reparametrizations of the integration variables. Upon introducing a set of auxiliary ``time'' variables, the Virasoro constraints can be compactly written as linear differential equations for the generating function \cite{Mironov:1990im}. Is is also often the case that this generating function of infinitely-many times acts as a $\tau$-function of some integrable hierarchy, thus satisfying an additional set of bilinear Hirota equations.
It is then clear that matrix models form a special class of solutions to both sets of equations: Virasoro and Hirota. However, given a generic solution to the Virasoro constraints it is not obvious (and in general not true) that this also solves Hirota bilinear equations. One can conclude that the moduli space of solutions of Virasoro constraints and that of Hirota relations intersect along a subspace containing the matrix model solutions but the precise relation between the two is still not completely understood (see \cite{Mironov:2019pij} for a review).

Inspired by this classical story of integrable models, we investigate such phenomena for generalizations of the Hermitean matrix model (HMM). Namely, we consider Virasoro constraints derived from HMMs with arbitrary polynomial potentials and we study their most general space of solutions. In particular, it is well known that in the case of a Gaussian potential the constraints can be solved combinatorially through repeated application of \textit{cut-and-join} operators to some specific initial data \cite{Morozov:2009xk}. In the case of higher polynomial potentials this technique can be generalized but the solution one finds is no longer unique, and one finds a larger space of possible initial data \cite{Cordova:2016jlu,Cassia:2020uxy}.

A secondary type of generalization which is natural to consider consists in deforming the domain of integration of the matrix model. In the HMM case one usually integrates over all Hermitean matrices, or equivalently over all possible values of the real eigenvalues (upon restriction of the integral to diagonal matrices). It is then interesting to ask what kind of Ward identities one can derive if the integral is restricted to a subspace of such matrices. For simplicity one can consider subspaces invariant under the adjoint action of $U(N)$ in such a way that the diagonalization procedure still applies and the integral can again be rewritten as an integral over eigenvalues. One of the simplest such deformations is the one in which we restrict each eigenvalue to some finite interval on the real line. This defines an integral over an $N$-dimensional box (hypercube) with boundary. This results in a boundary contribution to the Virasoro constraints, rendering them non-homogeneous. However, as we will show, the resulting constraints can still be solved.

The paper is organized as follows. In Section \ref{sec:homogeneous} we consider homogeneous Virasoro constraints, where we firstly review the Hermitean matrix model and the Virasoro constraints it satisfies. The most generic form of their solution is derived for potentials of low polynomial degrees. Secondly, we digress into the subject of classical integrability and integrable hierarchies of equations and investigate the conditions under which our solutions to the Virasoro constraints also satisfy Hirota bilinear equations. In Section \ref{sec:non-homogeneous} we move on to discuss non-homogeneous Virasoro constraints, where we provide the solution when the eigenvalue model is allowed to have a domain of integration with non-empty boundary. We discuss the specific examples in which the domain of integration is taken to be an hypercube or an orthant in $\mathbb{R}^N$. In Section \ref{sec:conclusion} we summarize our results and discuss further directions. In the Appendices we provide definitions for some special functions, we discuss the case of Virasoro constraints for ABJ-like matrix models and finally we provide a cohomological description of the constraints in terms of Lie algebra cohomology.

\section{Solving homogeneous Virasoro constraints}
\label{sec:homogeneous}

We begin by reviewing the classical Virasoro constraints satisfied by the Hermitean 1-matrix model, as inspired by \cite{Morozov:2009xk} in the case of a Gaussian potential and \cite{Cassia:2020uxy} in the case of a more general polynomial potential. We will here extend the results of \cite{Cassia:2020uxy} by providing additional details to the solutions of the Virasoro constraints, in particular by specifying normalisations (i.e. empty correlators) and also providing an explicit example for the case of cubic potential. 

We consider the $\beta$-deformed Hermitean matrix model given by the generating function 
\begin{equation}
\label{eq:beta_hermitean_mm_potential}
 \tau_N (\ak;t) = \frac{1}{N!}\int_{\Gamma^N}\,\prod_{i=1}^N \mathd x_i \prod_{1 \leq i < j \leq N}
 |x_i - x_j|^{2\beta} \mathe^{-\sum_{i=1}^N V(x_i) + \sum_{s=1}^{\infty}  t_s \sum_{i=1}^N x_i^s}
\end{equation}
where $N\in\mathbb{Z}_{\geq0}$ labels the rank of the matrices and $\beta \in \mathbb{C}$ is a generic deformation parameter \cite{Morozov:2012dz}. We take the potential $V(x)$ to be a polynomial function of the form
\begin{equation}
\label{eq:potential}
 V(x) = \sum_{k=1}^{\p} \frac{\ak_k x^k}{k}
\end{equation}
with coefficients or \textit{couplings} $\ak_k\in\mathbb{C}$.
The contour $\Gamma$ is chosen among all such that the integral is convergent as a functions of the parameters $\ak_k$. It is also important that all eigenvalues $x_i$ have the same integration contour $\Gamma$ so that the Weyl symmetry of the model is preserved, i.e. symmetry under permutations of the integration variables. This type of polynomial potential can equivalently be obtained as shift in the first $\p$ times
\begin{equation}
\label{eq:shiftoftimes}
 t_s \mapsto t_s - \ak_s/s, \quad\quad\quad s=1,\dots,\p~.
\end{equation}
In formula \eqref{eq:beta_hermitean_mm_potential} we use the condensed notations $\ak = \{\ak_1, \dots , \ak_{\p}\}$ to denote dependence on the polynomial coefficients and $t = \{t_1, t_2, \dots \}$ to denote dependence on an infinite set of formal {\it time variables}.
Here it should be noted that the infinite set of times $\{t_s \}$ are formal variables, in the sense that the power series expression of the generating function in these variables is not required to be convergent, only the coefficients in this expansion, i.e. the \textit{correlators}, are required to converge. On the other hand, these correlators are well defined analytical functions of the coupling constants $\{\ak_k\}$.

The model in \eqref{eq:beta_hermitean_mm_potential} satisfies the (homogeneous) Virasoro constraints
\begin{equation}
\label{eq:virasoro_constraints}
 \left(
 \sum_{k=1}^{\p} \ak_k \frac{\partial}{\partial t_{k+n}}
 + \delta_{n,-1} \ak_1 N
 - L_n\right) \tau_N (\ak;t) = 0 , \qquad n \geq - 1
\end{equation}
with the Virasoro generators $L_n$ being given by
\begin{equation}
\label{eq:virasoro_generators}
\begin{aligned}
L_{n>0} &= 2\beta N \frac{\partial}{\partial t_n} + \beta\sum_{\mu+\nu=n} \frac{\partial^2}{\partial t_\mu \partial t_{\nu}}+(1 - \beta) (n + 1) \frac{\partial}{\partial t_n} + \sum_{s>0} s t_s \frac{\partial}{\partial t_{s + n}} \\
L_0 &= \beta N^2 + (1-\beta)N+\sum_{s>0} s t_s \frac{\partial}{\partial t_{s }} \\
L_{-1} &= N t_1 +\sum_{s >0} s t_s \frac{\partial}{\partial t_{s-1}}~,
\end{aligned}
\end{equation}
In the literature it is often common to introduce an additional time $t_0$ and to let the generating function be proportional to $\mathe^{Nt_0}$ so that derivatives w.r.t. $t_0$ act as multiplications by factors of $N$.
From this point of view we can further write the generating function as a linear combination of functions at different values of $N$ without modifying the Virasoro constrains.
However, because the $t_0$-derivative operator commutes with all Virasoro generators, we can restrict ourselves to a given eigenspace with fixed eigenvalue $N$ without any loss of generality.
We also remark that from the point of view of the matrix model integral in \eqref{eq:beta_hermitean_mm_potential} the parameter $N$ is required to be an integer, however once we consider the Virasoro constraints in \eqref{eq:virasoro_constraints} this parameter does not need to satisfy any special condition and in fact it can be taken to be any complex number. As we will see in the end of this section, the solution to the constraints can always be analytically continued in $N$.

Because of \eqref{eq:shiftoftimes} we also have the additional constraints
\begin{equation}
\label{eq:additional_constraint}
 \left( \frac{\partial}{\partial t_k} + k \frac{\partial}{\partial \ak_k}\right) \tau_N (\ak;t) = 0\,,
 \qquad k = 1,\dots,\p
\end{equation}
whose solution can be trivially written as
\begin{equation}
 \tau_N (\ak;t) \equiv \tau_N (\{t_s-\ak_s/s\})
\end{equation}
We also remark that equations \eqref{eq:virasoro_constraints} and \eqref{eq:additional_constraint} are invariant under the action of the symmetry
\begin{equation}
\label{eq:betasymmetry}
\begin{aligned}
 \beta \to \frac{1}{\beta},
 \quad\quad\quad
 N \to -\beta N,
 \quad\quad\quad
 t_s \to -\frac{1}{\beta}t_s,
 \quad\quad\quad
 \ak_k \to -\frac{1}{\beta}\ak_k~.
\end{aligned}
\end{equation}
where we regard $\beta$ and $N$ as arbitrary complex parameters. Therefore we can expect the solution to satisfy the relation
\begin{equation}
\label{eq:tausymmetry}
 \tau_{N}^{\beta} (\ak;t) = \tau_{-\beta N}^{1/\beta} \left(-\frac{1}{\beta}\ak;-\frac{1}{\beta}t\right)
\end{equation}
for an appropriate choice of initial data on both sides.
In the context of AGT correspondence \cite{Alday:2009aq,Wyllard:2009hg} the parameter $\beta$ is the ratio of Nekrasov parameters $\epsilon_1$, $\epsilon_2$ and the symmetry $\beta\leftrightarrow 1/\beta$ corresponds to the exchange $\epsilon_1\leftrightarrow\epsilon_2$.

Starting from the Virasoro constraints in \eqref{eq:virasoro_constraints} one can recursively solve this in order to obtain correlators $c_\lambda(\ak)$ defined by the expansion
\begin{equation}
\label{eq:tau_expansion}
 \tau_N(\ak;t) = \sum_{\lambda}\frac{1}{|\Aut(\lambda)|} c_\lambda (\ak) \prod_{s\in\lambda}t_s~,
\end{equation}
as was shown in \cite{Cassia:2020uxy}, where the correlators $c_\lambda (\ak)$ are only determined up to an overall normalization factor corresponding to the empty correlator $c_\emptyset(\ak)$. Here the sum runs over all integer partitions $\lambda$, with $\Aut(\lambda)$ being the automorphism group of the partition, i.e. the group of permutations that leave the partition $\lambda$ invariant.

The case $\p=0$ is the most degenerate in the sense that the only solution is the constant generating function $\tau_N(t) \equiv \tau_N(0)$ and all correlators are identically zero. This is also reflected by the fact that the corresponding matrix model correlators are expressed by integrals which do not converge, due to the absence of an appropriate potential.

In the case of $\p=1,2$, it has been shown in \cite{Morozov:2009xk,Cassia:2020uxy} that the Virasoro constraints can be re-summed in order to express the generating function through cut-and-join operators.
We review here the derivation of such solutions. For $\p=1$ one can multiply each Virasoro constraint in \eqref{eq:virasoro_constraints} by a weight factor $(n+1) t_{n+1}$ and re-sum them all together from $n=0$ to $n=\infty$ to obtain a single equation
\begin{equation}
\label{eq:p1_virasoro_resummed}
 \ak_1 \, D \tau_N(\ak_1;t) = W_{-1} \tau_N(\ak_1;t)~.
\end{equation}
where $D$ is the dilatation operator written as
\begin{equation}
\label{eq:dilatation}
 D = \sum_{n=1}^\infty nt_n \frac{\partial}{\partial t_n}
\end{equation}
and $W_{-1}$ is the cut-and-join operator given by the series
\begin{equation}
 W_{-1} = \sum_{n=0}^\infty (n+1) t_{n+1}  L_{n}~.
\end{equation}
Observe that the dilatation operator $D$ acts diagonally on monomials with eigenvalues given by the total degree of the monomial~\footnote{We define a notion of $t$-degree such that $\deg(t_s)=s$ and $\deg(t_at_b)=\deg(t_a)+\deg(t_b)$. Then $D$ can equivalently be regarded as the corresponding grading operator.}.
Using the fact that $W_{-1}$ is homogeneous of degree 1 w.r.t. $D$, we can solve \eqref{eq:p1_virasoro_resummed} as follows. First we notice that $D$ annihilates the constant term in $\tau_N$. With some simple manipulations we can write
\begin{equation}
 \left(\ak_1 \, D-W_{-1}\right) \left[\tau_N(\ak_1;t)-\tau_N(\ak_1;0)\right] = W_{-1}\tau_N(\ak_1;0).
\end{equation}
Having removed the constant term from the generating function in the l.h.s., we now have that $D$ acts with non-zero eigenvalues and it becomes invertible
\begin{equation}
 \left(1 -\frac{1}{\ak_1}D^{-1}W_{-1}\right) \left[\tau_N(\ak_1;t)-\tau_N(\ak_1;0)\right] = \frac{1}{\ak_1}D^{-1}W_{-1}\tau_N(\ak_1;0)~.
\end{equation}
Finally we use the identity $\frac{1}{1-z}=\sum_{s=0}^\infty z^s$ to invert the operator on the l.h.s. and we get
\begin{equation}
\begin{aligned}
 \tau_N (\ak_1;t) &= \tau_N(\ak_1;0) + \sum_{s=0}^\infty \left(\frac{D^{-1}W_{-1}}{\ak_1}\right)^{s+1}\tau_N(\ak_1;0) \\
 &= \exp \left( \frac{W_{-1}}{\ak_1} \right) \tau_N(\ak_1;0)~.
\end{aligned}
\end{equation}
where in the last step we used that $\deg(W_{-1})=1$ so that all factors of $D^{-1}$ can be substituted with actual numbers that combine to give a factor $1/(s+1)!$ in the sum. Notice that all the time dependence of the generating function comes from the exponential of the cut-and-join operator, while $\tau_N(\ak_1;0)\equiv c_{\emptyset}(\ak_1)$ can be regarded as initial data.  

For $\p=2$ we can perform a similar computation to get the cut-and-join expression of the solution. In this case we re-sum the Virasoro constraints in \eqref{eq:virasoro_constraints} with weight $(n+2) t_{n+2}$ from $n=-1$ to $n= \infty$ and obtain
\begin{equation}
 \ak_2 \, D \, \tau_N(\ak_1,\ak_2;t) = \left( W_{-2} -\ak_1 \, L_{-1} \right) \tau_N(\ak_1,\ak_2;t) 
\end{equation}
where 
\begin{equation}\label{eq:W-2}
 W_{-2} = \sum_{n=-1}^\infty (n+2) t_{n+2} L_{n} ~,\quad\quad\quad \deg(W_{-2})=2~.
\end{equation}
In this case the cut-and-join operator in the r.h.s. is not homogeneous (in fact $\deg(L_{-1})=1$) however, using that $[W_{-2},L_{-1}]=0$, we can still get a compact formula for the solution,
\begin{equation}
\label{eq:WrepresentationP2}
 \tau_N(\ak_1,\ak_2;t) = \exp\left(\frac{W_{-2}}{2\ak_2}-\frac{\ak_1}{\ak_2}L_{-1}\right) c_{\emptyset}(\ak_1,\ak_2) ~.
\end{equation}
For $\ak_1=0$, $\ak_2=1$ and $\beta=1$ this matches exactly with the original solution of \cite{Morozov:2009xk}.

For $\p=3$ and higher we do encounter a problem. Namely, in order to get a dilatation operator in the l.h.s. of the equation we need to multiply by the weight $(n+\p)t_{n+\p}$ and then sum from $n+\p= 1$ to infinity, which means $n\geq 1-\p$. However we immediately see that for $n < -1$ there is no corresponding Virasoro constraint to sum. This clearly first happens when $\p=3$ in which case we would need to sum starting from the $L_{-2}$ constraint which is not defined. The best we can do in this case is to start the sum from the lowest constarint, i.e. $n=-1$. The resulting equation is
\begin{equation}
 \ak_\p \hat{D}\tau_N(\ak;t) = \hat{W}\tau_N(\ak;t)
\end{equation} 
with
\begin{equation}
\label{eq:partialdilatation}
 \hat{D}=\sum_{n=-1}^\infty (n+\p)t_{n+\p}\frac{\partial}{\partial t_{n+\p}}
\end{equation}
and
\begin{equation}
 \hat{W} = \sum_{n=-1}^\infty (n+\p) t_{n+\p} L_{n}
  - \sum_{k=1}^{\p-1} \ak_k \sum_{n=1-\p}^\infty (n+\p) t_{n+\p} \frac{\partial}{\partial t_{n+k}}
  - \ak_1(\p-1)t_{\p-1} N ~.
\end{equation}
Because $\p\geq3$ the sum in \eqref{eq:partialdilatation} is missing some of the initial terms, which means that the operator $\hat{D}$ has a larger kernel and it is no longer a good grading operator.
In particular, the operator $\hat{D}$ is blind to all times $t_s$ with $s<\p-1$.

Let us consider the example $\p=3$. The kernel of $\hat{D}$ in this case contains the constant term as well as all powers of $t_1$. As a consequence of this fact, all correlators of the form $c_{\{1,1,\dots,1\}}$ cannot be fixed by the recursion and they have to be regarded as additional initial data for the equation. For convenience we package all such missing data in a generating function of such correlators,
\begin{equation}
\label{eq:genfunt1}
\begin{aligned}
 \tau_N(\ak_1,\ak_2,\ak_3;t_1) &:= \left.\tau_N(\ak_1,\ak_2,\ak_3;t)\right|_{t_{s>1}=0} \\
  &= \sum_{k=0}^\infty c_{\{\underbrace{\scriptstyle1,\dots,1}_{k}\}}(\ak_1,\ak_2,\ak_3)\frac{t_1^k}{k!}
\end{aligned}
\end{equation}
in other words we drop all higher times from the generating function and we only keep the series in $t_1$.
We can now repeat the derivation of the previous cases and find the partial solution
\begin{equation}
 \tau_{N}(\ak_1,\ak_2,\ak_3;t) = \sum_{s=0}^{\infty}\left(\frac{\hat{D}^{-1}\hat{W}}{\ak_3}\right)^s
 \tau_{N}(\ak_1,\ak_2,\ak_3;t_1)
\end{equation}
Unfortunately, we can no longer give a closed formula for the series. First we observe that while $\hat{D}$ is still invertible on the image of $\hat{W}$, the cut-and-join operator $\hat{W}$ is a sum of non-commuting terms of different degrees with respect to $\hat{D}$. This implies that even if we substitute each factor of $\hat{D}^{-1}$ with its numerical value all such terms no longer combine to give a nice combinatorial factor. We are not aware of any nice resummation property of this series.

As we have shown in the previous paragraphs a solution to the Virasoro constraints \eqref{eq:virasoro_constraints} can always be found combinatorially provided a certain amount of initial data is provided. One should regard this data as boundary conditions to fix in order to solve the PDEs obtained from the constraints. The choice of such initial data however seems to be completely arbitrary. For some specific choice one recovers the generating function of the matrix model in \eqref{eq:beta_hermitean_mm_potential}, but for more general choices the solution the we constructed might not have a matrix integral representation.
Furthermore, there are cases in which the solution does admit a matrix model representation but not of the type of \eqref{eq:beta_hermitean_mm_potential}, for instance the ABJ-like model which we review in Appendix~\ref{sec:ABJ}.

The last step is to impose the additional equations \eqref{eq:additional_constraint}. This will restrict further the possible choices of initial data (as a function of the coupling constants).

\subsubsection{Degree $\p=1$}

The Virasoro constraints in \eqref{eq:virasoro_constraints} imply an infinite set of recursion relations between the coefficients of the power series of the generating function in \eqref{eq:tau_expansion}.
The additional constraint in \eqref{eq:additional_constraint} gives instead an infinite set of relations between correlators and their $\ak_1$-derivatives. Of all such relations we just need to consider the first non-trivial one, i.e. the one we get by setting all times to zero because all other relations will follow by our recursive solution of the Virasoro constraints. Hence we are led to consider the ODE
\begin{equation}
 \left. \frac{\partial}{\partial t_1} \tau_N (\ak_1;t)\right|_{t=0} = -\left. \frac{\partial}{\partial \ak_1}\tau_N (\ak_1;t)\right|_{t=0}
\end{equation}
where setting the times to zero corresponds to restricting to the initial data only.
Explicitly we can then write this equation as
\begin{equation}
\label{eq:c1fromadditional}
 c_{\{1\}} (\ak_1) = - \frac{\partial }{\partial \ak_1} c_\emptyset (\ak_1)~.
\end{equation}
But now we can use that the Virasoro constraints already tell us how to compute the correlator $c_{\{1\}}$,
\begin{equation}
\label{eq:c1fromvirasoro}
 c_{\{1\}}(\ak_1) = \frac{N(\beta(N-1)+1)}{\ak_1}\, c_\emptyset(\ak_1)~.
\end{equation}
Equating the r.h.s. of \eqref{eq:c1fromadditional} to that of \eqref{eq:c1fromvirasoro} we get a differential equation for $c_{\emptyset}$ as a function of $\ak_1$, which we can solve as
\begin{equation}
\label{eq:p1_solution}
 c_\emptyset (\ak_1) = k_{N,\beta}\cdot\ak_1^{-N(\beta(N-1)+1)}
\end{equation}
where $k_{N,\beta}$ is an integration constant which might depend on $N$ and $\beta$ but not on the coupling. This fixes the initial data essentially uniquely as a function of the coupling constant $\ak_1$ and therefore also the full generating function.

\subsubsection{Degree $\p=2$}

This case follows similarly. The additional constraint \eqref{eq:additional_constraint} now gives rise to 2 independent conditions, one for each coupling constant:
\begin{equation}
 c_{\{1\}}(\ak_1,\ak_2) = - \frac{\partial}{\partial\ak_1}c_\emptyset(\ak_1,\ak_2)~,
\end{equation}
\begin{equation}
 c_{\{2\}}(\ak_1,\ak_2) = -2 \frac{\partial}{\partial \ak_2} c_\emptyset (\ak_1,\ak_2)~.
\end{equation}
Substituting the correlators $c_{\{1\}}$, $c_{\{2\}}$ in the l.h.s. with the expressions we get by expanding \eqref{eq:WrepresentationP2},
\begin{equation}
 c_{\{1\}}(\ak_1,\ak_2) = -\frac{\ak_1 N}{\ak_2} \,c_\emptyset (\ak_1,\ak_2)~,
\end{equation}
\begin{equation}
 c_{\{2\}}(\ak_1,\ak_2) = \frac{N(\ak_2(\beta(N-1)+1)+\ak_1^2)}{\ak_2^2}\,c_\emptyset (\ak_1,\ak_2)~,
\end{equation}
we end up with 2 independent PDEs for the function $c_{\emptyset}(\ak_1,\ak_2)$. The solution can be computed exactly and it gives
\begin{equation}
\label{eq:p2_solution}
 c_\emptyset(\ak_1,\ak_2) = k_{N,\beta}\cdot \ak_2^{-\frac{1}{2}N(\beta(N-1)+1)}
 \exp\left(\frac{N\ak_1^2}{2\ak_2}\right)
\end{equation}
so that we can fix completely the dependence on the couplings up to some integration constant $k_{N,\beta}$.

\subsubsection{Degree $\p=3$}

In this case we have an unbalance between the number of coupling constants and that of non-trivial PDEs coming from the constraints. In particular, we observe that we can write 3 independent differential equations coming from the additional constraint \eqref{eq:additional_constraint}, however only two of them are meaningful once we substitute the solution of the Virasoro constraints. This is because the Virasoro recursion does not allow us to solve for all correlators. In other words we do not just have one function $c_{\emptyset}$ to determine but an infinite number of functions, i.e. all the coefficients of \eqref{eq:genfunt1}.

The additional constraint can be split into the three equations
\begin{equation}
\label{eq:p3_first}
 -\frac{\partial}{\partial \ak_1} c_\emptyset (\ak_1,\ak_2,\ak_3) = c_{\{1\}}(\ak_1,\ak_2,\ak_3)~,
\end{equation}
\begin{equation}
\label{eq:p3_second}
 -2 \frac{\partial}{\partial \ak_2} c_\emptyset (\ak_1,\ak_2,\ak_3) =
 \left(-\frac{\ak_1 N}{\ak_3} + \frac{\ak_2}{\ak_3}\frac{\partial}{\partial\ak_1} \right)
 c_\emptyset(\ak_1,\ak_2,\ak_3)~,
\end{equation}
\begin{equation}
\label{eq:p3_third}
 -3 \frac{\partial}{\partial \ak_3} c_{\emptyset}(\ak_1,\ak_2,\ak_3) =
 \left(\frac{(1-\beta (N-1))N}{\ak_3} - \frac{\ak_2^2}{\ak_3^2}\frac{\partial}{\partial\ak_1}
 + \frac{\ak_1 \ak_2 N}{\ak_3^2} +\frac{\ak_1}{\ak_3} \frac{\partial}{\partial \ak_1}\right)
 c_{\emptyset}(\ak_1,\ak_2,\ak_3)~,
\end{equation}
where in the first equation we are not able to substitute the explicit expression of $c_{\{1\}}$ as it cannot be determined using Virasoro. In fact, the correlator $c_{\{1\}}$ is itself part of the initial data. The other two equations however do contain non-trivial information and can be solved w.r.t. the couplings $\ak_2$ and $\ak_3$:
\begin{equation}
\label{eq:p3_solution}
 c_\emptyset (\ak_1,\ak_2,\ak_3) =
 \exp\left(-\frac{\ak_2 \left(\ak_2^2-6 \ak_1 \ak_3\right) N}{12 \ak_3^2}\right)
 \ak_3^{-\frac{1}{3}(1-\beta(N-1))N}
 g\left(\frac{\ak_2^2-4 \ak_1 \ak_3}{2 \ak_3^{4/3}}\right) ~.
\end{equation}
We observe that the solution is no longer unique up to normalization, and in fact it depends upon an arbitrary choice of function $g(z)$. Fixing this function explicitly then fixes completely the generating function in \eqref{eq:genfunt1} using the equation in \eqref{eq:p3_first}. To summarize, for $\p=3$ we cannot give a unique solution to the system of constraints however we managed to reduce all of the indeterminacy of the solution to a finite amount of information carried by the univariate function $g$. Moreover, $g$ might depend also on $N$ and $\beta$ but we do not write this dependence explicitly.

If we want to further restrict the solution then we need more conditions to impose on the function $g(z)$. For instance, if we insist that our solution is of matrix model type, then we need to impose other constraints that one might naturally obtain from matrix integrals. Namely, one can impose non-trivial relations between correlators that express the fact that traces of finite size matrices are not all independent. In fact, given a matrix $A$ of size $N$, only the first $N$ traces $\Tr(A^s)$, $s=1,\dots,N$ are linearly independent while all higher traces can be expressed as polynomial combinations of those\footnote{From the group theoretical point of view this corresponds to the fact that the algebra of $U(N)$ has $N$ linearly independent Casimir functions.}. In the language of the eigenvalue model, this means that all power-sum variables $p_s=\sum_{i}x_i^s$ of degree $s$ higher than the rank $N$ can be written in terms of $p_1,\dots,p_N$ and therefore similar relations must follow for the correlators. Let us consider the simple case of rank $N=1$ as an example to illustrate this point. For 1-dimensional matrices we have that the trace commutes with taking powers, therefore
\begin{equation}
 p_s = p_1^s~,
\end{equation}
and therefore two correlators must be equal if their partitions have the same size. It is straightforward then to obtain the corresponding Ward identity on the generating function $\tau_1(\ak;t)$,
\begin{equation}
\label{eq:p3generatingf}
 \frac{\partial}{\partial t_s}\tau_1(\ak_1,\ak_2,\ak_3;t)
  = \left(\frac{\partial}{\partial t_1}\right)^s\tau_1(\ak_1,\ak_2,\ak_3;t)~.
\end{equation}
The only independent equation is that for $s=2$ which combined with the additional constraint \eqref{eq:additional_constraint} becomes
\begin{equation}
 -2\frac{\partial}{\partial\ak_2}\tau_1(\ak_1,\ak_2,\ak_3;t)
  = \left(-\frac{\partial}{\partial \ak_1}\right)^2\tau_1(\ak_1,\ak_2,\ak_3;t)~.
\end{equation}
Applying to the solution in \eqref{eq:p3_solution} we finally find the equation
\begin{equation}
\label{eq:p3AiryN=1}
 z g(z) - 8 g''(z) = 0, \quad\quad\quad z=\frac{\ak_2^2-4 \ak_1 \ak_3}{2 \ak_3^{4/3}}
\end{equation}
which (up to rescaling of $z$) is the Airy ODE (see \eqref{eq:airyODE}), so that
\begin{equation}
 \left.g(z) \right|_{N=1} = k_A \AiryAi(z/2) + k_B \AiryBi(z/2)
\end{equation}
for arbitrary coefficients $k_A,k_B$. Similar considerations apply to rank $N>1$.

\subsubsection{Comments about degree $\p>3$}

The situation in degree $\p>3$ is similar to that of $\p=3$, in the sense that the solution is not unique and the initial data spans an infinite-dimensional space generated by all correlators whose partitions only contain integers that are smaller than $\p-2$. These are exactly the correlators that couple to monomials in times which are in the kernel of the partial dilatation operator $\hat{D}$ in \eqref{eq:partialdilatation}. Of the $\p$ additional contraints only $2$ can be used to obtain non-trivial equations for $c_\emptyset(\ak)$. These equations allow to reduce the number of functional variables of $c_\emptyset(\ak)$ from $\p$ to $\p-2$.

\subsection{Symmetry and universality of correlators}
\label{sec:universality}

Now that we found explicit solutions to the constraints we can go back to the symmetry equation in \eqref{eq:tausymmetry} and check explicitly that this is satisfied for our generating functions. We notice first that this is indeed the case when the solution is unique ($\p=1,2$) provided we rescale both sides of the equation \eqref{eq:tausymmetry} by the corresponding empty correlator, which is the only piece of initial data required. In the cases for which the solution is not unique ($\p\geq3$) then more assumptions on the initial data are required.

Assuming now that the initial data for the solution is chosen to be compatible with the symmetry $\beta\to1/\beta$, one can show that there is a special value of $\beta$ such that the correlators simplify and the solution becomes homogeneous in $N$, $\ak_k$ and $t_s$. Namely, for $\beta=-1/N$ the generating function satisfies the property
\begin{equation}
 \frac{\tau_{N}(\ak;t)}{\tau_{N}(\ak;0)}
 = \frac{\tau_{N/\kappa}(\kappa\,\ak;\kappa\,t)}{\tau_{N/\kappa}(\kappa\,\ak;0)}
 \quad\quad\quad
 \text{for }\beta=-\frac{1}{N}
\end{equation}
for an arbitrary scalar $\kappa\neq0$. This implies that $(N,\ak,t)$ are not independent parameters of the generating function. Using this property we can scale away the rank by choosing $\kappa=N$, hence we obtain the universal relation
\begin{equation}
 {\tau_{N}(\ak;t)}
 = \frac{\tau_{1}(N\ak;N t)}{\tau_{1}(N\ak;0)}{\tau_{N}(\ak;0)}
  \quad\quad\quad
 \text{for }\beta=-\frac{1}{N}
\end{equation}
which expresses the rank $N$ generating function as a scalar multiple of the rank $1$ (normalized) generating function. As an immediate consequence of this fact we have that all correlators in rank $N$ become universal in the sense that they only depend on the size of the corresponding partition and not on the actual shape (exactly as in the rank 1 case). Moreover, these correlators do depend on the rank $N$ as simple functions, namely their dependence only enters through a power $N^{\ell(\lambda)}$, where $\ell(\lambda)$ is the length of the partition.

\subsection{A digression into Hirota equations}
\label{subsec:Hirota}

In this section we analyze whether our solutions of Virasoro constraints are actual $\tau$-functions, by which we mean generating functions in times which satisfy a bilinear equation analogous to the Pl\"ucker identities for a finite dimensional Grassmannian. This integrability condition is called Hirota equation and, following work of Sato, it can be regarded as the infinite dimensional analogue of Pl\"ucker relations for the Sato Grassmannian (for a review see references \cite{Kac:1601455,1791585,Alexandrov_2013}).

Hirota bilinear equations are written as the integral identity
\begin{equation}
\label{eq:hirota3}
 \oint{\mathrm{d}z}
 \exp\left(2\sum_{s=1}^\infty z^{s}v_s\right)
 \exp\left(-\sum_{s=1}^\infty\frac{z^{-s}}{s}\frac{\partial}{\partial v_s}\right)\tau(u+v)\tau(u-v) = 0
\end{equation}
where $u_s$, $v_s$ are two independent sets of times, $z$ is an auxiliary complex variable and $\tau$ is a generic generating function. The integration in $z$ is used to project out the coefficient of $z^{-1}$ in the Laurent series expansion of the integrand. If this residue is zero then we say that $\tau$ is a $\tau$-function.

Equation \eqref{eq:hirota3} is a rather compact way to express Hirota relation but not very explicit in terms of what it implies for the correlators. In order to get a better understanding of its meaning we can expand the generating function $\tau$ as a power series in the times $v_s$ as
\begin{equation}
 \tau(u\pm v) = \sum_{d=0}^\infty [\tau(u\pm v)]_{(d)}
 = \sum_{d=0}^\infty\left[\sum_{\lambda\vdash d} \frac{1}{|\mathrm{Aut}(\lambda)|}
\tau_{\lambda}(u) \prod_{l\in\lambda}\left( \pm v_{l}\right) \right] ~,
\end{equation}
where $[\tau(u \pm v)]_{(d)}$ is used to denote the part of $\tau(u \pm v)$ which is of degree $d$ in the times $v_s$ and $\lambda\vdash d$ denotes that $\lambda$ is an integer partition of $d\in\mathbb{Z}_{\geq0}$. Here we also use the notation $\tau_\lambda (u) = \prod_{a \in \lambda} \frac{\partial}{\partial u_a} \tau(u)$. Similarly, we expand the exponentials using Cauchy's identity
\begin{equation}
 \exp\left(\sum_{s=1}^\infty \frac{z^s p_s}{s}\right)
 = \sum_{m=0}^\infty z^m \schur_{\{m\}}(p_s)~,
\end{equation}
where $\schur_{\{m\}}(p_s)$ is the symmetric Schur polynomial of order $m$. We can therefore rewrite \eqref{eq:hirota3} as
\begin{equation}
\label{eq:Hirota}
\begin{aligned}
 & \frac{1}{2\pi\mathi}\oint{\mathrm{d}z} \sum_{k=0}^\infty\sum_{l=0}^\infty z^{k-l}
 \mathsf{Schur}_{\{k\}}\left(p_s=2sv_s\right)
 \mathsf{Schur}_{\{l\}}\left(p_s=-\frac{\partial}{\partial v_s}\right)
 \tau(u+v)\tau(u-v) = \\
 & = \sum_{k=0}^\infty
 \mathsf{Schur}_{\{k\}}\left(p_s=2sv_s\right)
 \mathsf{Schur}_{\{k+1\}}\left(p_s=-\frac{\partial}{\partial v_s}\right)
 \tau(u+v)\tau(u-v) = \\
 & = \sum_{d=0}^{\infty} \left[ \sum_{k=0}^{d}
 \mathsf{Schur}_{\{k\}}\left(p_s=2sv_s\right)
 \mathsf{Schur}_{\{k+1\}}\left(p_s=-\frac{\partial}{\partial v_s}\right)
 \sum_{j=0}^{d+1}[\tau(u+v)]_{(j)}[\tau(u-v)]_{(d+1-j)}\right] = 0~.
 \end{aligned}
\end{equation}
The expression inside the square brackets in the last line represents the projection of \eqref{eq:hirota3} onto its degree $d$ part w.r.t. the times $v_s$. Since each degree $d$ must be independent from the others, we get an infinite set of relations. Moreover, by explicit computation one can check that for degree $0\leq d < 3$ this equation is trivially satisfied, while the first non-trivial condition on the $\tau_\lambda(u)$ appears at degree $d=3$.

In order to get an equation for the correlators we need to set the times $u_s$ to zero so that
\begin{equation}
 \tau_\lambda (0) \equiv c_{\lambda}~.
\end{equation}
For instance, in degree $d=3$ we get the identity
\begin{equation}
\label{eq:Hirota_3}
\begin{split}
 3 c _{\{1,1\}}^2 &+3 c _{\emptyset} c _{\{2,2\}}-4 c _{\emptyset} c _{\{3,1\}}-4 c _{\{1\}} c _{\{1,1,1\}}+ c _{\emptyset} c _{\{1,1,1,1\}}-3 c _{\{2\}}^2+4 c_{\{1\}} c _{\{3\}} = 0 ~,
\end{split}
\end{equation}
and in degree $d=4$
\begin{equation}
\label{eq:Hirota_4}
 3 c _{\{1,1\}} c _{\{2,1\}}+2 c_{\emptyset} c _{\{3,2\}}-3 c _{\{1\}} c _{\{2,1,1\}}+c _{\emptyset} c _{\{2,1,1,1\}}+
  3 c _{\{1\}} c_{\{4\}} -3 c _{\emptyset} c _{\{4,1\}}-c_{\{2\}} c _{\{1,1,1\}}-2 c _{\{2\}} c _{\{3\}} =0 ~.
\end{equation}
In degree 5 and higher there are more than one independent bilinear equation (one for each partition of size $d$).

Matrix models (at $\beta=1$) are known to provide solutions to Hirota equations and the reason is that the generating function can be written as a determinant of a $N\times N$ symmetric matrix (known as the Hankel matrix). This property follows from Andr\'eief's integration formula \cite{An86} and it is known to fail when a $\beta$-deformation is turned on. Restricting to the case $\beta=1$, Andr\'eief's integration formula allows us to write the determinantal identity
\begin{equation}
\label{eq:det_rep}
 \tau_N(\ak;t) = \underset{1\leq i,j \leq N}{\det}\left[\left(\frac{\partial}{\partial t_1}\right)^{i+j-2} 
 \tau_1(\ak;t)\right]~.
\end{equation}
This way we are able to express the rank $N$ matrix integral as the determinant of a Hankel matrix constructed out of simpler rank 1 integrals,
\begin{equation}
 \tau_1(\ak;t) = \int_{\Gamma} \mathd x \, \mathe^{- V(x) + \sum_{s=1}^\infty t_s x^s}~.
\end{equation}
Now, assuming the form of the potential $V(x)$ given in \eqref{eq:potential}, one can trade the derivatives with respect to $t_1$ with derivatives in the coupling constant $\ak_1$ using the additional constraint in \eqref{eq:additional_constraint}. Evaluating the generating function at zero times then gives a similar determinant expression for the empty correlator of the matrix model (i.e. the partition function)
\begin{equation}
\label{eq:det_rep_empty}
 \tau_N(\ak;0) = \underset{1 \leq i,j \leq N}{\det}\left[ \left(-\frac{\partial}{\partial\ak_1} \right)^{i+j-2} \tau_1(\ak;0)  \right]
\end{equation}
For arbitrary values of the $\beta$-parameter there are no known generalizations of Hirota bilinear equations. However, for $\beta=1$ the matrix model generating functions do satisfy Hirota equations by construction and formula \eqref{eq:det_rep} gives a combinatorial way to compute all correlators in arbitrary finite rank. In the following we try to match the Virasoro solutions that we found in the previous section against explicit matrix model expressions. Generically speaking, the matching boils down to the choice of initial data for the recursion. Initial data compatible with matrix model integrals will automatically lead to Virasoro solutions which are also solutions to Hirota equations. More general choices of initial data, when plugged into Hirota equations give rise to additional non-trivial differential equations that one needs to impose.
For the sake of concreteness we explicitly consider the cases $\p=1, 2, 3$ and analyze them one by one.

\subsubsection{$\p=1$}

Because the solution in \eqref{eq:p1_solution} allows us to rewrite all correlators as scalar multiples of the empty one, once we plug them into the bilinear equation \eqref{eq:Hirota} there is an overall factor of $c_\emptyset^2$ which factors out of the equation. This means that Hirota equations are blind to the choice of initial data for the $\p=1$ Virasoro solution. Since the overall normalization of the correlators is no longer important we can rescale $c_\emptyset$ to match with that of a matrix model with linear potential. Hence the matrix integral solution is essentially the unique one. For arbitrary $\beta$ this solution does not satisfy Hirota however when $\beta=1$ it does automatically. This can be checked using the rank $N=1$ integral
\begin{equation}
 \tau_1(\ak_1;0) = \int_0^\infty \mathe^{-\ak_1 x} \mathd x = \frac{1}{\ak_1},
 \quad\quad\quad \Re(\ak_1)>0
\end{equation}
together with the determinantal formula \eqref{eq:det_rep_empty} which becomes
\begin{equation}
\begin{aligned}
 \tau_N(\ak_1;0) & = \underset{1\leq i, j \leq N}{\det} \left[ \Gamma(i+j-1) \ak_1^{1-i-j}\right] \\
 & = \BarnesG(N+1)^2 \cdot \ak_1^{-N^2}~.
\end{aligned}
\end{equation}
This reproduces exactly the solution \eqref{eq:p1_solution} at $\beta=1$ once we appropriately choose the value of the integration constant $k_{N,1}=\BarnesG(N+1)^2$.

\subsubsection{$\p=2$}

A similar discussion holds for the case $\p=2$. The solution is also essentially unique and can be rescaled to match the matrix model one. At $\beta\neq1$ Hirota does not hold but for $\beta=1$ we can use Andr\'eief's integration formula to express the matrix model generating function as a determinant. An explicit computation gives
\begin{equation}
 \tau_1(\ak_1,\ak_2;0) = \int_{-\infty}^\infty \mathe^{-\ak_1 x - \ak_2 \frac{x^2}{2} } \mathd x
 = \sqrt{\frac{2\pi}{\ak_2}} \mathe^{\frac{\ak_1^2}{2 \ak_2}},
 \quad\quad\quad \Re(\ak_2)>0
\end{equation}
so that
\begin{equation}
\begin{aligned}
 \tau_N(\ak_1, \ak_2;0) & = \underset{1\leq i, j \leq N}{\det} \left[
 \left(- \frac{\partial}{\partial a_1}\right) ^{i+j-2} \sqrt{\frac{2\pi}{\ak_2}} \mathe^{\frac{\ak_1^2}{2 \ak_2}}
 \right] \\
 & = (2\pi)^{\frac{N}{2}} \BarnesG(N+1) \cdot \ak_2^{-\frac{N^2}{2}} \mathe^{\frac{N\ak_1^2}{2 \ak_2}}
\end{aligned}
\end{equation}
This reproduces the solution \eqref{eq:p2_solution} at $\beta=1$ for an appropriate choice of integration constant $k_{N,1}=(2\pi)^{\frac{N}{2}}\BarnesG(N+1)$.

\subsubsection{$\p=3$}

When the potential is of degree higher than 2 the correlators are not all proportional to the empty one and the actual Virasoro solution depends on more initial data. For $\p=3$ all correlators can be rewritten as polynomials in the totally anti-symmetric correlators contained in the generating function \eqref{eq:p3generatingf}. Using the additional constraint in \eqref{eq:additional_constraint} we can rewrite each one of them using $\ak$-derivatives acting on the empty correlator as
\begin{equation}
 c_{\{\underbrace{\scriptstyle1,\dots,1}_{k}\}}(\ak) = \left(-\frac{\partial}{\partial\ak_1}\right)^k c_\emptyset(\ak) ~.
\end{equation}
If we then plug this Virasoro solution inside of Hirota equations we get non-linear differential equations for $c_{\emptyset}(\ak)$. In degree $d=3$ for instance, equation \eqref{eq:Hirota_3} becomes
\begin{equation}
\begin{aligned}
 \ak_3^2
 \left(
  c_\emptyset \partial_{\ak_1}^4 c_\emptyset
  - 4 \left(\partial_{\ak_1}^3 c_\emptyset\right) \left(\partial_{\ak_1} c_\emptyset\right)
  + 3 \left(\partial_{\ak_1}^2 c_\emptyset \right)^2
 \right) +
 (\ak_2^2-4 \ak_1 \ak_3)
 & \left(
  \left(\partial_{\ak_1} c_\emptyset\right)^2 - c_\emptyset \partial_{\ak_1}^2 c_\emptyset
 \right) + \\
 & - 2 \ak_3 c_\emptyset \partial_{\ak_1} c_\emptyset
 + \ak_2 N c_\emptyset^2 = 0
\end{aligned}
\end{equation}
which is clearly a non-trivial constraint on the function $c_\emptyset(\ak)$ and it means that Hirota equations
are no longer automatically satisfied not even in the $\beta=1$ case. Using the explicit expression of $c_\emptyset$ that we found in \eqref{eq:p3_solution} we then get an equation for the unknown function $g$, namely
\begin{equation}
\label{eq:p3hirotad3forg}
 2 \left(z g'(z)^2-8 g'''(z) g'(z)+6 g''(z)^2\right)+g(z) \left(g'(z)-2 z g''(z)+4 g''''(z)\right) =0
\end{equation}
Again, this is a non-trivial differential constraint on the initial data $g(z)$ and therefore we have that some of our Virasoro solutions do not satisfy Hirota\footnote{Here we only considered the $d=3$ Hirota equation, however one should check for all equations in all degrees $d>3$.} even for $\beta=1$.

Clearly, there is one obvious set of solutions which is provided by matrix models at $\beta=1$, however in this case it is not so simple to write down a closed formula for the function $g(z)$ by matching with the integral expression of the generating function (especially at higher rank $N>1$). As an example we consider the simple case of $N=1$. The cubic potential $V(x)$ can always be rewritten as a depressed cubic via a change of variable $x=(\ak_3^{-1/3}x'-\ak_2/2\ak_3)$, so that we can write the rank-1 matrix integral as
\begin{equation}
\begin{aligned}
 \tau_1(\ak_1,\ak_2,\ak_3;0)
 & = \int_\Gamma \exp\left(-\frac{1}{3}\ak_3 x^3 - \frac{1}{2}\ak_2 x^2 - \ak_1 x\right) \mathd x \\
 & = \ak_3^{-1/3}\exp\left(-\frac{\ak_2(\ak_2^2-6\ak_1\ak_3)}{12\ak_3^2}\right)
 \int_{\Gamma'}\exp\left(-\frac{x'^3}{3}+\left(\frac{\ak_2^2-4\ak_1\ak_3}{4\ak_3^{4/3}}\right)x'\right)\mathd x'~.
\end{aligned}
\end{equation}
By comparison with \eqref{eq:p3_solution} we have
\begin{equation}
 \left.g(z)\right|_{N=1} = \int_{\Gamma'}\exp\left(-\frac{x'^3}{3}+\frac{z}{2}x'\right)\mathd x'~,
\end{equation}
which, for an appropriate choice of contour $\Gamma'$, is a linear combination of Airy functions.
It is then straightforward to show that \eqref{eq:p3hirotad3forg} and all higher degree Hirota equations follow from the Airy differential equation \eqref{eq:p3AiryN=1}. We should mention however, that it is not obvious to us that the converse is true as well, i.e. that Hirota equations of the form \eqref{eq:p3hirotad3forg} imply the Airy ODE on $g(z)$, which would imply that solution to Hirota plus Virasoro at $\p=3$ are essentially uniquely given by a 1-matrix integrals (at least for rank 1).

\section{Solving non-homogeneous Virasoro constraints}
\label{sec:non-homogeneous}

Now we wish to generalize the above matrix model computations and allow for the domain of integration of the eigenvalues to have a boundary. As a concrete example we focus on the case in which each eigenvalue $x_i$ belongs to the finite interval $[a,b]$ on the real line\footnote{Note here that $a$ is the limit of integration which is not to be confused with the couplings $\ak_k$ appearing in the potential $V(x)$ in \eqref{eq:potential}.}. The generating function of the matrix model is then given by
\begin{equation}
\label{eq:tau_boundary_general}
 \tau_N(\ak;t) = \frac{1}{N!}\int_{[a,b]^N} \prod_{i=1}^N \mathd x_i \prod_{1 \leq i < j \leq N}
 |x_i-x_j|^{2\beta} \mathe^{-\sum_{i=1}^N V(x_i) + \sum_{s=1}^\infty t_s \sum_{i=1}^N x_i^s}
\end{equation}
where the compact domain $[a,b]^N$ is essentially an hypercube inside of $\mathbb{R}^N$.
Furthermore, it is important that all integrals have the same domain of integration $[a,b]$ because we want to preserve the symmetry of the integral under permutation of the eigenvalues (this is the action of the Weyl group of $U(N)$ on the Cartan subalgebra). The choice of domain becomes part of the parameters of the generating function $\tau_N(\ak;t)$, however we do not explicitly write this dependence in order not to clutter too much the notation.

The matrix model in \eqref{eq:tau_boundary_general} no longer satisfies standard Virasoro constraints because none of the infinitesimal deformations generated by the vector fields
\begin{equation}
 \xi_n = \sum_{i=1}^N x_i^{n+1}\frac{\partial}{\partial x_i}, \quad\quad\quad n\geq-1
\end{equation}
preserve the domain of integration $[a,b]^N$. This means that the integral is no longer invariant under such deformations. However, we can compute exactly what is the variation of the integral and write down a modified version of the Virasoro constraints. As it turns out, these equations are modified by a boundary term which can be rewritten as a matrix integral of rank $N-1$. This boundary term can be directly associated to a matrix model living on the $(N-1)$-dimensional faces of the hypercube and it turns the Virasoro constraints into non-homogeneous equations.
More concretely, we can write the equations
\begin{equation}
\label{eq:inhomogeneousVirasoro}
\left( \sum_{k=1}^{\p}\ak_k \frac{\partial}{\partial t_{k+n} }+\ak_1 N \delta_{n,-1}
 -L_n \right) \tau_N (\ak;t) = B_n(t) , \qquad n \geq -1
\end{equation}
where the l.h.s. is the same as that of the standard Virasoro constraints, while the r.h.s. of the equation is the boundary term mentioned above.
As before, the generating function satisfies the additional constraints
\begin{equation}
\left( \frac{\partial}{\partial t_k} + k \frac{\partial}{\partial \ak_k}\right) \tau_N (\ak;t)= 0\, , \qquad k = 1, \dots \p
\end{equation}
due to the presence of a non-trivial polynomial potential.

We now compute each such boundary term $B_n(t)$ explicitly. Let us call $\omega(t)\in\Omega^N(\mathbb{R}^N)\otimes\mathbb{C}[[t_1,t_2,\dots]]$ the top degree differential form in the integrand of \eqref{eq:tau_boundary_general}, so that $\tau_N(t)=\int\omega(t)$.
Virasoro constraints are obtained by integrating the variation of $\omega(t)$ with respect to the infinitesimal diffeomorphism generated by the vector field $\xi_n$. We can therefore write this variation as a Lie derivative along the vector which becomes
\begin{equation}
 B_n(t) = -\int_{\mathcal{D}}\mathcal{L}_{\xi_n}\omega(t)
 = -\int_{\mathcal{D}} \mathd \iota_{\xi_n} \omega(t)
 = -\int_{\partial\mathcal{D}} \iota_{\xi_n} \omega(t)
\end{equation}
where $\mathcal{D}\subset\mathbb{R}^N$ is a generic domain of integration and we used that $\omega(t)$ is closed\footnote{The form $\mathd\omega$ might have some legs in the $t$-directions however these are projected out when contracted with $\xi_n$.} because of dimensional reasons.
Substituting the explicit expressions for $\omega(t)$, $\xi_n$ and $\mathcal{D}$ we have
\begin{equation}
 \iota_{\xi_n} \omega(t) =
 \frac{1}{N!} \sum_{i=1}^N (-1)^{i+1} x_i^{n+1} \prod_{j\neq i} \mathd x_i \prod_{1 \leq j < k \leq N}
  |x_j-x_k|^{2\beta} \mathe^{-\sum_{j=1}^N V(x_j) + \sum_{s=1}^\infty t_s \sum_{j=1}^N x_j^s}
\end{equation}
and integrating over $\partial[a,b]^N$ we get
\begin{equation}
\label{eq:boundaryterm}
 B_n(t) = -\left\llangle \prod_{j=1}^{N-1}|z-x_j|^{2\beta} \right\rrangle_{N-1}
 z^{n+1} \left.\exp\left(-V(z) + \sum_{s=1}^\infty t_s z^s\right)\right|_{z=a}^{z=b}
\end{equation}
where we introduced the notation
\begin{equation}
 \left.f(z)\right|_{z=a}^{z=b} = f(b)-f(a)
\end{equation}
and
\begin{equation}
 \left\llangle f(x) \right\rrangle_{N} :=
 \frac{1}{N!} \int_{[a,b]^{N}} \prod_{i=1}^{N}\mathd x_i\,
 f(x) \prod_{1 \leq i < j \leq N} |x_i-x_j|^{2\beta}
 \exp\left({-\sum_{i=1}^{N} V(x_i) + \sum_{s=1}^\infty t_s \sum_{i=1}^{N} x_i^s}\right)
\end{equation}
so that $\llangle\,\rrangle_N$ is the \textit{time-dependent} expectation value in a matrix model of rank $N$.

Before discussing the solution of these non-homogeneous Virasoro constraints we remark here that the choice of contour $[a,b]^N$ has a direct interpretation in terms of matrices and not just eigenvalues. Namely, one can define the function
\begin{equation}
 \sigma : H_N \to \mathbb{R}^N/S_N
\end{equation}
which assigns to each Hermitean matrix in $H_N$ its spectrum which can be represented as a $N$-tuple of real numbers up to permutations (hence the quotient by the symmetric group $S_N$). This map is surjective and moreover it is clearly invariant under the adjoint action of $U(N)$ on $H_N$. This means that the preimage of any subset of the codomain is a disjoint union of adjoint orbits of $U(N)$. The last step consists of identifying the appropriate subset of the codomain. We choose the compact subspace $[a,b]^N/S_N\hookrightarrow\mathbb{R}^N/S_N$ so that this choice matches the domain in \eqref{eq:tau_boundary_general} (up to a numerical factor equal to the order of the Weyl group). The preimage under $\sigma$ of this region defines a subset of $H_N$ invariant under the $U(N)$ action. This is the domain of integration of the actual matrix model.

\subsection{Gaussian constraints in the hypercube}

For the sake of concreteness, we now consider the particular case of a Gaussian potential with $\p=2$ and coupling constants $\ak_1=0$ and $\ak_2=1$, in other words $V_{\mathrm{G}}(x) = \frac{x^2}{2}$. Since the coupling constants are fixed, we avoid writing them as variables of the generating function and simply denote it as $\tau_N(t)$. Other choices of polynomial potentials are not conceptually more difficult to treat except for a slightly more involved discussion of the initial data. Here we chose a Gaussian potential so that we can demonstrate the main ideas for solving non-homogeneous constraints without having to discuss the possible non-uniqueness of the solution.

The non-homogeneous Gaussian Virasoro constraints are given by 
\begin{equation}
\label{eq:Virasoro_constraints_Gaussian}
 \left(\frac{\partial}{\partial t_{n+2}} -L_n \right) \tau_N(t) = B_n(t) ~, \quad\quad\quad n\geq-1~,
\end{equation}
with the boundary term $B_n(t)$ as in \eqref{eq:boundaryterm}.
Re-summing the constraints with weight $(n+2)t_{n+2}$ from $n = -1$ to $n= \infty$ we find the equation
\begin{equation}
 \left(D-W_{-2}\right) \tau_N(t) = B(t) 
\end{equation}
where $D$ and $W_{-2}$ are given in \eqref{eq:dilatation} and \eqref{eq:W-2} respectively, while
\begin{equation}
 B(t) := \sum_{n=-1}^\infty (n+2)t_{n+2} B_n(t) ~.
\end{equation}
This can be solved inductively on the degree as
\begin{equation}
 \tau_N^{(d)}(t) = \frac{1}{d}\left(W_{-2} \tau_N^{(d-2)}(t) +B^{(d)}(t) \right)
\end{equation}
where $d$ is the degree with respect to the dilatation operator $D$ and we assume the expansion 
\begin{equation}
\label{eq:B_expansion}
 B(t) = \sum_{d=1}^\infty B^{(d)}(t)~,\quad\quad\quad [D,B^{(d)}(t)] = d \cdot B^{(d)}(t) ~.
\end{equation}
Observe that the sum over $d$ starts from one and not from zero because from \eqref{eq:boundaryterm} it follows that $B^{(0)}(t)=B(0)=0$.
Then we have 
\begin{equation}
\label{eq:tau_boundary}
\begin{split}
 \tau_N(t) = & \tau_N(0) +(D-W_{-2})^{-1} (W_{-2}\tau_N(0) +B(t)) \\
 = & \sum_{s=0}^\infty (D^{-1}W_{-2})^s \tau_N(0) + \sum_{s=0}^\infty (D^{-1}W_{-2})^s D^{-1} B(t) \\
 = & \exp\left(W_{-2}/2 \right) \tau_N(0) + \sum_{s=0}^\infty (D^{-1}W_{-2})^s D^{-1} B(t) \\
 = & \exp(W_{-2}/2) \tau_N(0) + \sum_{s=0}^\infty \sum_{d=1}^\infty \frac{W_{-2}^s B^{(d)} (t) }{2^s d \left( \frac{d}{2}+1\right) _s} ~,
\end{split}
\end{equation}
with $(x)_n$ being the Pochhammer symbol defined in \eqref{eq:pochhammer}. The result above generalizes the solution to the homogeneous Virasoro constraints given in \eqref{eq:WrepresentationP2} for $\ak_k=\delta_{k,2}$.
Observe that the non-homogeneous solution naturally factorizes into the sum of a homogeneous solution and a boundary term depending on $B(t)$. See Appendix~\ref{sec:algebraic-description} for a cohomological interpretation of this fact.
 
In the following we provide explicit solutions to the non-homogeneous constraints in terms of correlators for the cases of low rank $N=1,2$ as examples. By doing so we also elaborate on the fact that the rank $N$ solution can be quite explicitly constructed from knowledge of the one in rank $N-1$, so that there are two types of recursions at play. The first is the usual recursion on correlators for fixed rank and the second is the recursion in the rank itself.

\subsubsection{Examples}

Starting with the case of $N=1$, the boundary term in \eqref{eq:boundaryterm} becomes
\begin{equation}
B_n(t) = a^{n+1} \exp\left( -\frac{a^2}{2} +\sum_{s=1}^\infty t_s a^s \right) - b^{n+1} \exp\left( -\frac{b^2}{2} +\sum_{s=1}^\infty t_s b^s \right) 
\end{equation}
so that the degree $d$ term in the expansion \eqref{eq:B_expansion} can be written as
\begin{equation}
 B^{(d)} (t) = \sum_{n=-1}^{d-2} (n+2)t_{n+2} \sum_{\lambda\vdash (d-n-2)} \frac{1}{| \textrm{Aut}(\lambda ) |} \left( a^{d-1} \mathe^{-\frac{a^2}{2}}-b^{d-1} \mathe^{-\frac{b^2}{2}} \right) \prod_{\ell \in \lambda} t_\ell ~.
\end{equation}
Plugging in \eqref{eq:tau_boundary} we can explicitly compute the first few correlators%
\footnote{Notice that in rank $N=1$ the correlators only depend on the size of the partitions and not on the actual shape.}
\begin{equation}
\begin{split}
 c^{N=1}_\emptyset & = \tau_1(0) \\
 c^{N=1}_{\{1\}} & = \mathe^{-\frac{a^2}{2}} - \mathe^{-\frac{b^2}{2}} \\
 c^{N=1}_{\{2\}} & = \tau_1(0) + a \mathe^{-\frac{a^2}{2}} - b \mathe^{-\frac{b^2}{2}} \\
 c^{N=1}_{\{3\}} & = \left(a^2+2\right) \mathe^{-\frac{a^2}{2}}-\left(b^2+2\right) \mathe^{-\frac{b^2}{2}} \\
 c^{N=1}_{\{4\}} & = 3\tau_1(0) + a\left(a^2+3\right)\mathe^{-\frac{a^2}{2}} - b\left(b^2+3\right)\mathe^{-\frac{b^2}{2}} \\
 c^{N=1}_{\{5\}} & = \left(a^4+4 a^2+8\right) \mathe^{-\frac{a^2}{2}}-\left(b^4+4 b^2+8\right) \mathe^{-\frac{b^2}{2}}~.
\end{split}
\end{equation}
The empty correlator in this case can be computed explicitly by evaluating the integral and we get
\begin{equation}
 \tau_1(0) = \int_{a}^{b}\exp(-x^2/2)\mathd x
 = \sqrt{\frac{\pi }{2}} \left(\erf\left(\frac{b}{\sqrt{2}}\right)-\erf\left(\frac{a}{\sqrt{2}}\right)\right)~,
\end{equation}
then we can write a closed formula for all correlators in rank 1 as
\begin{equation}
\label{eq:boundary_generic_correlator}
  c^{N=1}_{\lambda} = F_{|\lambda|}
\end{equation}
with
\begin{equation}
\begin{aligned}
  F_s := & \frac{(1+(-1)^{s})}{2}(s-1)!!\,\tau_1(0) +\\
  & + \left(\sum_{k=0}^{\left\lfloor\frac{s-1}{2}\right\rfloor}
  \frac{(s-1)!!}{(s-1-2k)!!}a^{s-1-2k}
  \right)\mathe^{-\frac{a^2}{2}}
  - \left(\sum_{k=0}^{\left\lfloor\frac{s-1}{2}\right\rfloor}
  \frac{(s-1)!!}{(s-1-2k)!!}b^{s-1-2k}
  \right)\mathe^{-\frac{b^2}{2}} ~.
\end{aligned}
\end{equation}
Moving on to the case of rank $N=2$ the boundary term takes the form
\begin{equation}
 B_n(t) = \left\llangle|a-x_1|^{2\beta} \right\rrangle_1 a^{n+1} \exp\left(-\frac{a^2}{2}
  + \sum_{s=1}^\infty t_s a^s \right)
  - \left\llangle|b-x_1|^{2\beta} \right\rrangle_1 b^{n+1} \exp \left( -\frac{b^2}{2} + \sum_{s=1}^\infty t_s b^s\right) ~.
\end{equation}
For arbitrary values of $\beta$ the expectation values that appear are not polynomial in the integration variables of the matrix model therefore this boundary term is a complicated function for which we do not have a closed formula. For integer $\beta$ however they become expectation values of polynomial functions and thus they can be computed combinatorially. For instance, let us consider the simple case $\beta=1$.
Then we can write the boundary term of degree $d$ as
\begin{equation}
\begin{aligned}
 B^{(d)} & (t) = \sum_{n=-1}^{d-2} (n+2)t_{n+2} \sum_{\substack{\lambda, \rho \\ |\lambda |+|\rho| = d-(n+2)}}
 \frac{1}{|\Aut(\lambda)|} \frac{1}{|\Aut(\rho)|} \prod_{\ell\in\lambda} t_\ell \prod_{r\in\rho} t_r \times \\
 & \times \left[ \left(F_{2 +|\lambda |} -2 a F_{1 +|\lambda |} + a^2 F_{|\lambda|} \right) a^{n+1+|\rho|} \mathe^{-\frac{a^2}{2}}
  - \left(F_{2+|\lambda |} -2 b F_{1 +|\lambda |} + b^2 F_{|\lambda|}\right) b^{n+1+|\rho|}\mathe^{-\frac{b^2}{2}} \right]  ~.
\end{aligned}
\end{equation}
We can now use \eqref{eq:tau_boundary} to explicitly find the first few correlators as
\begin{equation}
\begin{aligned}
 c^{N=2}_\emptyset &= \tau_2(0) \\
 c^{N=2}_{\{1\}} &= \mathe^{-\frac{a^2}{2}} \left(a^2 F_0 -2a F_1+F_2\right)
  - \mathe^{-\frac{b^2}{2}} \left(b^2 F_0 - 2b F_1+F_2\right)\\
 c^{N=2}_{\{1,1\}} &= 2 \tau_2(0) + \mathe^{-\frac{a^2}{2}} \left(a^3 F_0-a^2 F_1-a F_2+F_3\right)
  + \mathe^{-\frac{b^2}{2}} \left(b^3 F_0-b^2 F_1-b F_2+F_3\right)\\
 c^{N=2}_{\{2\}} &= 4 \tau_2(0) + a \mathe^{-\frac{a^2}{2}} \left(F_0 a^2-2 F_1 a+F_2\right)
  - b \mathe^{-\frac{b^2}{2}} \left(F_0 b^2-2 F_1 b+F_2\right) \\
  \dots &
\end{aligned}
\end{equation}
Similarly, one can use the data from rank 1 and 2 to construct the rank 3 solution and higher.

\subsection{Gaussian constraints in the orthant}

We now consider the special case of a domain $\mathcal{D}=[0,\infty)^N$, i.e. the orthant obtained as the limit in which $a\to0$ and $b\to\infty$. The generating function on the orthant is given by
\begin{equation}
\label{eq:tau_function_orth}
 \tau_N(t) = \frac{1}{N!}\int_{[0,\infty)^N} \prod_{i=1}^N \mathd x_i \prod_{1 \leq k < l \leq N} |x_k-x_l|^{2\beta} \mathe^{-\sum_{i=1}^N V_{\mathrm{G}}(x_i) + \sum_{s=1}^\infty t_s \sum_{i=1}^N x_i^s}
\end{equation}
where we again have a Gaussian potential $V_{\mathrm{G}}(x) = \frac{x^2}{2}$.
Since one of the boundaries has been pushed to infinity we are left with only one boundary contribution per integration variable, namely the contribution coming from the boundary at zero. By explicitly computing formula \eqref{eq:boundaryterm} we immediately see that all such terms vanish if $n\geq0$, while for $n=-1$ we have the only non-trivial boundary term
\begin{equation}
\label{eq:boundaryorthant}
 B_{n}(t) = \delta_{n,-1} \left\llangle \prod_{i=1}^{N-1}x_i^{2\beta} \right\rrangle_{N-1}~.
\end{equation}
Therefore the Virasoro constraints are almost homogeneous with the exception of the $n=-1$ constraint, also known as the {\it string equation} \cite{Dijkgraaf:1990rs}.
This suggests that we can use the following trick. We compute the solution to the homogeneous $n\geq0$ constraints independently of the string equation and then solve the string equation in a second step. In order to perform the first step, we resum all homogeneous constraints with weight $(n+2)t_{n+2}$ to get
\begin{equation}
\label{eq:resumVirorth}
 \hat{D}\tau_N(t) = \hat{W}_{-2}\tau_N 
\end{equation}
with
\begin{equation}
 \hat{D}=\sum_{s=2}^\infty st_s\partial_s
 \quad\quad\quad\text{and}\quad\quad\quad
 \hat{W}_{-2} = \sum_{n=0}^\infty (n+2)t_{n+2}L_{n}~.
\end{equation}
Observe that $\hat{D}$ is the same partial dilatation operator that we encountered in the cubic case $\p=3$ in \eqref{eq:partialdilatation}. This means that the solution of \eqref{eq:resumVirorth} is not unique and it must depend on a choice of initial data given by generating function of anti-symmetric correlators, i.e. $\tau_N(t_1)$ as defined in \eqref{eq:genfunt1}. The solution of the recursion then can be written as
\begin{equation}
\label{eq:partialsolution}
 \tau_N(t) = \sum_{s=0}^\infty (\hat{D}^{-1}\hat{W}_{-2})^s\, \tau_N(t_1)
\end{equation}
This fixes all correlators except for those of the form $c_{\{1,1,\dots\}}$ which can only be fixed in a second step by analyzing the string equation. Equivalently, using \eqref{eq:partialsolution} we can fix completely the dependence of the generating function upon all times $t_s$ with $s>1$. The dependence on the time $t_1$ is entirely contained in the function $\tau_N(t_1)$ and can only be fixed via the string equation. Setting all the higher times to zero we can simplify the string equation as
\begin{equation}
 \left(\partial_1-N t_1\right) \tau_N (t_1) = B_{-1}(t_1)~.
\end{equation}
We define the operator on the l.h.s. as $M_N:=\partial_1-N t_1$. Since $M_N$ is invertible on the complement of the homogeneous solution (i.e. the kernel of $M_N$), the solution becomes
\begin{equation}
 \tau_N (t_1) = \tau_N^{\mathrm{h}} (t_1) + M_N^{-1}B_{-1}(t_1)
\end{equation}
with $\tau_N^{\mathrm{h}}=\mathe^{Nt_1^2/2}\tau_{N}(0)$ being a solution to the homogeneous constraints.
Furthermore, we can study the boundary term $B_{-1}(t_1)$ and observe that for integer $\beta$ the boundary term is the expectation value of a polynomial function and therefore it can be rewritten as the action of a differential operator on the generating function in rank one less. We call this operator $S_{N-1}(t_1,\partial_1)$ and define it by the equation
\begin{equation}
 B_{-1}(t_1) =: S_{N-1}\, \tau_{N-1}(t_1)~.
\end{equation}
Finally, we can write the solution to the non-homogeneous string equation as
\begin{equation}
 \tau_{N}(t_1) = \mathe^{\frac{Nt_1^2}{2}}\tau_{N}(0) + M_N^{-1}S_{N-1}\tau_{N-1}(t_1)~,
\end{equation}
whose computation can now be done inductively w.r.t. the rank $N$.
We remark here that the operator $S_{N}$ is well-defined only for integer $\beta$ when the argument of the expectation value in \eqref{eq:boundaryorthant} is polynomial. For arbitrary $\beta$ one should compute the non-homogeneous contribution as in \eqref{eq:tau_boundary}.

For $\beta=1$ one can compute the operator $S_N$ for the first few values of $N$ by making use of the Virasoro constraints. For instance,
\begin{equation}
 S_{0}(t_1,\partial_1) = 1
\end{equation}
\begin{equation}
 S_{1}(t_1,\partial_1) = \partial_1^2
\end{equation}
\begin{equation}
 S_{2}(t_1,\partial_1) = \frac{1}{4} \left( \partial_1^4-12 \partial_1^2-2t_1\partial_1^3+11t_1 \partial_1+t_1^2 \partial_1^2+24\right)
\end{equation}
\begin{equation}
\begin{split}
 S_{3}(t_1,\partial_1) = & \frac{1}{36} (
104 t_1^3 \partial_1
+4 t_1^4 \partial_1^2
-264 t_1^2 \partial_1^2
-12 t_1^3 \partial_1^3
+13 t_1^2 \partial_1^4
+576 t_1^2\\
&
-1260 t_1 \partial_1
+639 \partial_1^2
+213 t_1 \partial_1^3
-6 t_1\partial_1^5
-54 \partial_1^4
+ \partial_1^6
-540)
\end{split}
\end{equation}
The claim is thus that using the above procedure the model can be solved completely in any rank, up to initial data given by the collection of all empty correlators $\tau_N({0})$.

\subsubsection{Examples}

Let us consider first the case of rank $N=1$. The boundary terms in the Virasoro constraints on the orthant are
\begin{equation}
 B_n(t) = \delta_{n,-1}
 \quad\quad\quad
 B(t) = t_1
\end{equation}
Plugging in \eqref{eq:tau_boundary} we get
\begin{equation}
\label{eq:exampleN1orthant}
 \tau_1(t) = \exp\left(W_{-2}/2\right)\tau_1(0)
 + \sum_{s=0}^\infty \frac{W_{-2}^s }{2^s(3/2)_s}t_1~.
\end{equation}
The string equation is
\begin{equation}
 (\partial_1-t_1)\tau_1(t_1) = 1
\end{equation}
whose solution is
\begin{equation}
\label{eq:exampleN1orthantsol}
 \tau_1(t_1) = \mathe^{\frac{t_1^2}{2}} \tau_1(0)
  + \sqrt{\frac{\pi }{2}} \mathe^{\frac{t_1^2}{2}} \erf\left(\frac{t_1}{\sqrt{2}}\right)
\end{equation}
which is exactly what we get from \eqref{eq:exampleN1orthant} upon setting all higher times to zero (so that $W_{-2}=t_1^2$).
Here $\tau(0)$ is an integration constant which we can fix by computing the following integral explicitly
\begin{equation}
 \tau_1(0) = \int_{0}^{\infty} \exp(-x^2/2) \mathd x = \sqrt{\frac{\pi}{2}}~,
\end{equation}
which we can use to give a closed formula for each correlator as
\begin{equation}
\label{eq:n=1sol}
 c_{\lambda}^{N=1} = 2^{\frac{|\lambda|-1}{2}} \Gamma \left(\frac{|\lambda|+1}{2}\right)~.
\end{equation}
This is consistent with \eqref{eq:boundary_generic_correlator} upon taking the limit $a \rightarrow 0$ and $b \rightarrow \infty$.

For rank $N=2$ we have
\begin{equation}
 B_{-1}(t) = \left\llangle x^{2\beta} \right\rrangle_{1}
 = \int_{0}^\infty x^{2\beta}\exp\left(-\frac{x^2}{2}+\sum_{s=1}^\infty t_s x^s\right)\mathd x
\end{equation}
and
\begin{equation}
 B^{(d)}(t) = \sum_{|\lambda|=d-1} \frac{1}{|\mathrm{Aut}(\lambda)|} 2^{\frac{|\lambda|-1+2\beta}{2}} \Gamma \left(\frac{|\lambda|+1+2\beta}{2}\right) t_1 \prod_{a\in\lambda} t_a ~.
\end{equation}
In this case we cannot get a closed formula for the correlators, however we can still get explicit expressions for individual correlators by using \eqref{eq:tau_boundary}. The first few correlators are
\begin{equation}
\begin{aligned}
c^{N=2}_{\emptyset} & = \tau_2(0) \\
c^{N=2}_{\{1\}} & = 2^{\beta-\frac{1}{2}} \Gamma \left(\beta+\frac{1}{2}\right) \\
c^{N=2}_{\{1,1\}} & = 2\tau_2(0) + 2^{\beta} \Gamma (\beta+1)  \\
c^{N=2}_{\{2\}} & = 2 (\beta+1) \tau_2(0) \\
c^{N=2}_{\{1,1,1\}} & = 2^{\beta-\frac{1}{2}} (2 \beta+5) \Gamma \left(\beta+\frac{1}{2}\right) \\
c^{N=2}_{\{2,1\}} & = \frac{2^{\beta+\frac{3}{2}} \Gamma \left(\beta+\frac{5}{2}\right)}{2 \beta+1}\\
c^{N=2}_{\{3\}} & = 2^{\beta+\frac{1}{2}} (\beta+1) \Gamma \left(\beta+\frac{1}{2}\right) ~.
\end{aligned}
\end{equation}
In general, one can compute the rank $N$ solution using knowledge of the rank $N-1$ correlators.

\section{Conclusion}
\label{sec:conclusion}

In the present article we studied various versions of Virasoro constraints obtained as Ward identities for Hermitean 1-matrix models with polynomial potential. First we analyzed the moduli space of solutions for the standard Virasoro constraints and compared the solution on a generic point in this space to that given by the Hermitean matrix model itself. In this regard we showed that there must exist points in the moduli space which do not satisfy Hirota bilinear relations and we investigated the minimal set of assumptions to define the subspace of Hirota solutions.

An interesting question that arises then is whether there are solutions to Hirota that do not come from solving Virasoro constraints. One possible set of candidates for such $\tau$-functions are those obtained from $q,t$-models at $t=q$ (or $\beta=1$) for which the matrix integral is substituted by a Jackson integral. These model satisfy $q$-Virasoro constraints instead of usual ones, however they do still admit a determinant representation via Andr\'eief's identity and therefore they satisfy the Hirota equation. These and other discrete generalizations of the Hermitean matrix models certainly do merit further examination.

Next, we studied non-homogeneous Virasoro constraint obtained from certain matrix models whose integration domains has non-empty boundary. We showed that a solution can be algebraically constructed in a similar way as in the homogeneous case and we further argued that such solutions can be obtained by induction on the rank of the model.

It would be interesting to study more general cases of non-homogeneous Virasoro constraints. In particular, one could consider domains of integration defined by inequalities of the type
\begin{equation}
 \label{eq:Casimirsubspace}
 p_s = \sum_{i=1}^N x_i^s \leq C
\end{equation}
for some fixed constant $C\in\mathbb{R}$ (which generalizes the role of our $a$ and $b$ parameters). This leads to a well defined matrix model because the power-sums are Casimirs of the Lie algebra of $U(N)$ hence they define conjugation invariant functions on the space of Hermitean matrices. Moreover, each Virasoro vector field $\xi_n$ is proportional to the gradient of the power-sum function $p_{n+2}$ which means that every subspace defined by \eqref{eq:Casimirsubspace} is maximally deformed by one of the Virasoro generators. Correspondingly, at least one constraints will be non-homogenous and the boundary term should have some geometric interpretation in terms of the theory of coadjoint orbits for $U(N)$. We leave the study of these models for future investigations.

\appendix

\section{Special functions}

In this section we will summarize the special functions used throughout. The first such function is the Gamma function $\Gamma(z)$ defined as the analytic continuation of the integral
\begin{equation}
\label{eq:Gamma_def}
 \Gamma(z) =
 \int_0^{\infty} t^{z-1} \mathe^{-t} \mathd t \,, \qquad \Re(z)>0
\end{equation}
from which we can define the Pochhammer symbol $(z)_n$,
\begin{equation}
\label{eq:pochhammer}
 (z)_n = \frac{\Gamma (z+n)}{\Gamma (z) }~.
\end{equation}
We also need the Airy functions which are defined as being the linearly independent solutions to the differential equation
\begin{equation}
\label{eq:airyODE}
 f''(z) -z f(z) = 0
\end{equation}
i.e. 
\begin{equation}
 f(z) = k_A \, \AiryAi(z) + k_B \, \AiryBi(z)~.
\end{equation}
with $k_A,k_B$ arbitrary complex coefficients. 
The Airy functions can be given in terms of the contour integrals
\begin{equation}
\label{eq:airy_a}
 \AiryAi(z) = \frac{1}{2\pi\mathi} \int_{\infty\mathe^{-\frac{\pi\mathi}{3}}}^{\infty\mathe^{\frac{\pi  \mathi}{3}}} \exp\left(\frac{t^3}{3}-t z\right) \, \mathd t
\end{equation}
and
\begin{equation}
\label{eq:airy_b}
 \AiryBi(z) = \frac{1}{2\pi}\int_{-\infty}^{\infty\mathe^{-\frac{\pi\mathi}{3}}}
 \exp\left(\frac{t^3}{3}-t z\right) \, \mathd t \,
 + \, \frac{1}{2\pi}\int_{-\infty }^{\infty\mathe^{\frac{\pi\mathi}{3}}}
 \exp\left(\frac{t^3}{3}-t z\right) \, \mathd t ~.
\end{equation}
Furthermore, we make use of the error function $\erf(z)$ which we define by the series
\begin{equation}
\label{eq:error_function}
 \erf(z) = \frac{2}{\sqrt{\pi}}\sum_{n=0}^{\infty}\frac{(-1)^nz^{2n+1}}{n!(2n+1)} ~,
\end{equation}
and the Barnes G-function
\begin{equation}
\label{eq:barnesG_function}
 \BarnesG(z+1)=(2\pi )^{z/2}\exp \left(-{\frac {z+z^{2}(1+\gamma )}{2}}\right)\,\prod _{k=1}^{\infty }\left\{\left(1+{\frac {z}{k}}\right)^{k}\exp \left({\frac {z^{2}}{2k}}-z\right)\right\} ~,
\end{equation}
for $\gamma$ the Euler-Mascheroni constant. For positive integers $n\in\mathbb{Z}$ the Barnes G-function evaluates to
\begin{equation}
 \BarnesG(n+1) = \prod_{k=1}^{n-1} k!~.
\end{equation}

\section{ABJ-like models}
\label{sec:ABJ}

We consider here a matrix model of non-Hermitean type which does satisfy the same Virasoro (homogeneous) constraints as those considered in the bulk of this article. Let us define a $\beta$-deformed ABJ-like matrix model by the matrix integral generating function 
\begin{equation}
\label{eq:betaABJ}
\begin{aligned}
 \tau_{N,M} (t) = \frac{1}{N!}\frac{1}{M!}
 \int_{\Gamma_x^N}\,\prod_{i=1}^N \mathd x_i \int_{\Gamma_y^M}\,\prod_{a=1}^M \mathd y_a &
 \frac{
 \prod_{1 \leq i < j \leq N} |x_i - x_j|^{2\beta}
 \prod_{1 \leq a < b \leq M} |y_a - y_b|^{2/\beta}
 }{\prod_{i=1}^N\prod_{a=1}^M |x_i - y_a|^{2}} \times\\
 &\times  \exp\left(
 \sum_{s=1}^{\infty}  t_s \left(\sum_{i=1}^N x_i^s-\frac{1}{\beta}\sum_{a=1}^M y_a^s\right)\right)
\end{aligned}
\end{equation}
where we observe that there are two sets of eigenvalues variables $\{x_i\}$ and $\{y_a\}$, each integrated over the corresponding contour $\Gamma_x$ or $\Gamma_y$, respectively. Moreover, the familiar Vandermonde determinant has been replaced by a ($\beta$-deformed) Cauchy determinant in the first line. This model can also be interpreted as the matrix model corresponding to an integral over supermatrices in the algebra of the supergroup $U(N|M)$ \cite{Drukker:2009hy} which plays an important role in computing the localized partition function of the ABJ theory \cite{Aharony:2008ug,Aharony:2008gk} and also for the analytically continued Chern-Simons partition function on lens space \cite{Marino:2009jd}.

The expression in \eqref{eq:betaABJ} is to be regarded as a formal integral due to the absence of a potential which would make the integral convergent, however we remark that a polynomial potential can always be introduced by shifting the time variables as in \eqref{eq:shiftoftimes}.
The Virasoro constraints can then be derived by acting with the Lie derivative along the vectors
\begin{equation}
 \xi_n = \sum_{i=1}^N x_i^{n+1}\frac{\partial}{\partial x_i}
  + \sum_{a=1}^M y_a^{n+1}\frac{\partial}{\partial y_a}
\end{equation}
and they can be rewritten as
\begin{equation}
\label{eq:ABJconstr}
 \left( \sum_{k=1}^{\p} \ak_k\frac{\partial}{\partial t_{k+n}}
 + \delta_{n,-1}\ak_{1}\left(N-\frac{1}{\beta}M\right) - L^\mathrm{ABJ}_n\right)\tau_{N,M}(\ak;t) = 0~,
 \quad\quad\quad n\geq-1
\end{equation}
with operators $L_n^{\mathrm{ABJ}}$ given by 
\begin{equation}
\begin{aligned}
 L^{\mathrm{ABJ}}_{n>0} &= 2\beta\left(N-\frac{1}{\beta}M\right)\frac{\partial}{\partial t_n}
 + \beta\sum_{\mu+\nu=n}\frac{\partial^2}{\partial t_\mu \partial t_\nu}
 + \left(1-\beta\right) (n + 1) \frac{\partial}{\partial t_n}
 + \sum_{s>0} s t_s \frac{\partial}{\partial t_{s + n}} \\
 L^{\mathrm{ABJ}}_0 &= \beta\left(N-\frac{1}{\beta}M\right)^2 + \left(1-\beta\right)\left(N-\frac{1}{\beta}M\right)
 + \sum_{s>0} s t_s \frac{\partial}{\partial t_{s }} \\
 L^{\mathrm{ABJ}}_{-1} &= \left(N-\frac{1}{\beta}M\right) t_1 +\sum_{s >0} s t_s \frac{\partial}{\partial t_{s -1}}~.
\end{aligned}
\end{equation}
It is then clear that if we define the effective rank
\begin{equation}
 N_{\mathrm{eff}} := N-\frac{1}{\beta}M
\end{equation}
the constraints in \eqref{eq:ABJconstr} became equivalent to those in \eqref{eq:virasoro_constraints} for the HMM. Therefore, for an appropriate choice of initial data, the solutions to the two sets of equations will be equivalent. In other words,
\begin{equation}
 \tau_{N,M}\left(\ak;t\right) \cong \tau_{N_\mathrm{eff}}\left(\ak;t\right)
\end{equation}
This shows that the same Virasoro constraints do have different solutions from the HMM one, in this case not of matrix model type but supermatrix model.

Finally, we observe that the ABJ generating function in \eqref{eq:betaABJ} and the constraints in \eqref{eq:ABJconstr} are both invariant under the symmetry transformation
\begin{equation}
 \beta \to \frac{1}{\beta},
 \quad\quad\quad
 N_{\mathrm{eff}} \to -\beta N_{\mathrm{eff}},
 \quad\quad\quad
 t_s \to -\frac{1}{\beta}t_s,
 \quad\quad\quad
 x_i \leftrightarrow y_a,
 \quad\quad\quad
 N \leftrightarrow M.
\end{equation}
Notice that this is a generalization of the symmetry in \eqref{eq:betasymmetry} which can easily be recovered for $M=0$. Moreover, there is a special value of $\beta$ for which the solution becomes homogeneous w.r.t. a rescaling of effective rank and times, namely for
\begin{equation}
 \beta = -\frac{1-M}{N}
\end{equation}
we have that the generating function satisfies
\begin{equation}
 \frac{\tau_{N,M}(\ak;t)}{\tau_{N,M}(\ak;0)} = \frac{\tau_{N/\kappa,M}(\kappa\ak;\kappa\,t)}{\tau_{N/\kappa,M}(\kappa\ak;0)}
 = \frac{\tau_{N,\kappa(M-1)+1}(\kappa\ak;\kappa\,t)}{\tau_{N,\kappa(M-1)+1}(\kappa\ak;0)}~,
 \quad\quad\quad \kappa\in\mathbb{C}^\times
\end{equation}
which then implies the universal relation
\begin{equation}
 \tau_{N,M}(\ak;t) = \frac{\tau_{1,0}\left(\frac{N}{1-M}\ak;\frac{N}{1-M}t\right)}{\tau_{1,0}\left(\frac{N}{1-M}\ak;0\right)} \tau_{N,M}(\ak;0)~.
\end{equation}

\section{Cohomological description of the constraints}
\label{sec:algebraic-description}

In this section we use Lie algebra cohomology (with coefficients in a module) to describe the relation between homogeneous and non-homogeneous Virasoro constraints.
In particular we restrict to the Gaussian case, i.e. $\p=2$ and $\ak_1=0$, $\ak_2=1$.

We consider the parabolic subalgebra $\mathfrak{p}\subset\mathrm{Vir}$
generated by $\{U_{-1},U_{0},U_{1},\dots\}$, with
$U_n := L_n-\partial_{n+2}$.
Let $V_N=\mathbb{C}[[t_1,t_2,\dots]]$ be
the ring of formal series in times together with a
$\mathfrak{p}$-module structure given by the action of the Virasoro
generators as differential operators in times. Observe that all $V_N$ for $N\in\mathbb{Z}$ are isomorphic to each other as vector spaces but not as $\mathfrak{p}$-modules. They differ by the action of the Virasoro generators $U_n$ which explicitly depends on the integer $N$.

We define cochains
$C^n(\mathfrak{p},V_N)\cong \mathrm{Hom}(\Lambda^n\mathfrak{p},V_N)$ and a
differential $\delta:C^n(\mathfrak{p},V_N)\to C^{n+1}(\mathfrak{p},V_N)$,
\begin{equation}
\begin{aligned}
 (\delta f)(X_1,\dots,X_{n+1}) =& \sum_{1\leq s<t\leq n+1} (-1)^{s+t-1} f([X_s,X_t],X_1,\dots,\hat{X}_s,\dots,\hat{X}_t,\dots,X_{n+1}) \\
 & +\sum_{s=1}^{n+1} (-1)^s X_s \cdot f(X_1,\dots,\hat{X}_s,\dots,X_{n+1})
\end{aligned}
\end{equation} with $f\in C^n(\mathfrak{p},V_N)$ and
$X_s\in\mathfrak{p}$.

A 1-cochain $\alpha$ is a 1-cocycle if it is $\delta$-closed, i.e.
\begin{equation}
 \delta \alpha(X_1,X_2) = -\alpha([X_1,X_2]) - X_1\cdot\alpha(X_2) + X_2\cdot\alpha(X_1) =0
\end{equation} and it is exact if it is the $\delta$-image of a
0-cochain $f\in C^0(\mathfrak{g},V_N)\cong V_N$, \begin{equation}
 \alpha(X) = \delta f (X) = -X\cdot f
\end{equation}
If we define a 1-cochain
$\alpha\in C^1(\mathfrak{g},V_N)$ by $\alpha(U_n) := B_n$, with $B_n\in V_N$ defined as in \eqref{eq:boundaryterm}, then we can interpret
equation \eqref{eq:inhomogeneousVirasoro} as saying that
$\tau_N\in V_N$ is a 0-cochain and that $\alpha$ is its
differential,
\begin{equation}
 \delta \tau_N = \alpha
\end{equation}
so that $\alpha$ is trivial in cohomology and $\tau_N$ is a trivialization.
In the Lie algebra cohomology language one says that $\alpha$ is an inner derivation of $\mathfrak{p}$.
Moreover, two different trivializations differ by a 0-cocycle $Z_N\in H^0(\mathfrak{p},V_N)\subset V_N$, i.e. a solution to the homogeneous Virasoro constraints
\begin{equation}
 \delta Z_N(U_n) = -U_n \cdot Z_N = 0, \quad\quad\quad n\geq-1~.
\end{equation}
Therefore we have the familiar statement that non-homogeneous solutions form an affine space modeled over the space of homogeneous solutions  $H^0(\mathfrak{p},V_N)$.

Because $\alpha$ is exact then it must also be closed and its cocycle condition can be written as
\begin{equation}
 \delta\alpha (U_m,U_n) = -\alpha([U_m,U_n]) - U_m \cdot \alpha(U_n) + U_n \cdot \alpha(U_m) =0~,
\end{equation}
which we can rewrite using the explicit expression of the commutator in $\mathfrak{p}$ to get
\begin{equation}
\label{eq:cocycle}
 (m-n) B_{m+n} = -U_m \cdot B_n + U_n \cdot B_m~.
\end{equation}
This means that \eqref{eq:cocycle} is a necessary (but not sufficient) condition on the non-homogeneous terms $B_n$ in order for the constraints to have a solution.

\providecommand{\href}[2]{#2}\begingroup\raggedright\endgroup

\end{document}